\documentclass[12pt]{article}

\def\ben{\begin{equation}}
\def\een{\end{equation}}

  \let\n=\nu

\let\C=\Chi

\def\nn{\nonumber} \def\bd{\begin{document}} \def\ed{\end{document}}
\def\ds{\documentstyle} \let\fr=\frac \let\bl=\bigl \let\br=\bigr
\let\Br=\Bigr \let\Bl=\Bigl
\let\bm=\bibitem
\let\na=\nabla
\let\pa=\partial \let\ov=\overline
\newcommand{\be}{\begin{equation}}
\newcommand{\ee}{\end{equation}}
\def\ba{\begin{array}}
\def\ea{\end{array}}
\def\ft#1#2{{\textstyle{{\scriptstyle #1}\over {\scriptstyle #2}}}}
\def\fft#1#2{{#1 \over #2}}
\def\del{\partial}
\def\vp{\varphi}
\def\sst#1{{\scriptscriptstyle #1}}
\def\oneone{\rlap 1\mkern4mu{\rm l}}
\def\td{\tilde}
\def\wtd{\widetilde}
\def\ie{\rm i.e.\ }
\def\dalemb#1#2{{\vbox{\hrule height .#2pt
        \hbox{\vrule width.#2pt height#1pt \kern#1pt
                \vrule width.#2pt}
        \hrule height.#2pt}}}
\def\square{\mathord{\dalemb{6.8}{7}\hbox{\hskip1pt}}}
\newcommand{\ho}[1]{$\, ^{#1}$}
\newcommand{\hoch}[1]{$\, ^{#1}$}
\newcommand{\bea}{\begin{eqnarray}}
\newcommand{\eea}{\end{eqnarray}}
\newcommand{\ra}{\rightarrow}
\newcommand{\lra}{\longrightarrow}
\newcommand{\Lra}{\Leftrightarrow}
\newcommand{\ap}{\alpha^\prime}
\newcommand{\bp}{\tilde \beta^\prime}
\newcommand{\tr}{{\rm tr} }
\newcommand{\Tr}{{\rm Tr} }
\def\0{{\sst{(0)}}}
\def\1{{\sst{(1)}}}
\def\2{{\sst{(2)}}}
\def\3{{\sst{(3)}}}
\def\4{{\sst{(4)}}}
\def\5{{\sst{(5)}}}
\def\6{{\sst{(6)}}}
\def\7{{\sst{(7)}}}
\def\8{{\sst{(8)}}}
\def\n{{\sst{(n)}}}
\def\cA{{{\cal A}}}
\def\cB{{{\cal B}}}
\def\cF{{{\cal F}}}
\def\tV{\widetilde V}
\def\tW{\widetilde W}
\def\tH{\widetilde H}
\def\tE{\widetilde E}
\def\tF{\widetilde F}
\def\tA{\widetilde A}
\def\im{{{\rm i}}}
\def\tY{{{\wtd Y}}}
\def\ep{{\epsilon}}
\def\vep{{\varepsilon}}
\def\R{\rlap{\rm I}\mkern3mu{\rm R}}
\def\bD{{{\bar D}}}

\def\R{\rlap{\rm I}\mkern3mu{\rm R}}
\def\bD{{{\bar D}}}
\def\R{{{\Bbb R}}}
\def\C{{{\Bbb C}}}
\def\H{{{\Bbb H}}}
\def\CP{{{\Bbb C}{\Bbb P}}}
\def\RP{{{\Bbb R}{\Bbb P}}}
\def\Z{{{\Bbb Z}}}
\def\bA{{{\Bbb A}}}
\def\bB{{{\Bbb B}}}
\def\bC{{{\Bbb C}}}
\def\bD{{{\Bbb D}}}
\def\bE{{{\Bbb E}}}
\def\bZ{{{\Bbb Z}}}
\def\Re{{{\frak{Re}}}}
\def\Im{{{\frak{Im}}}}
\def\cosec{{\,\hbox{cosec}\,}}
\def\Gm{{\Gamma_{\!\! -}}}
\def\Gp{{\Gamma_{\!\! +}}}
\def\stan{{standard }}
\def\nonstan{{supernumerary }}

\newcommand{\tamphys}{\it Center for Theoretical Physics,
Texas A\&M University, College Station, TX 77843}

\newcommand{\upenn}{\it Department of Physics and Astronomy,\\ University
of Pennsylvania, Philadelphia, PA 19104}

\newcommand{\brussels}{\it Physique Th\'eorique et Math\'ematique,
Universit\'e Libre de Bruxelles,\\ Campus Plaine C.P. 231, B-1050
Bruxelles, Belgium}

\newcommand{\auth}{H. L\"u\hoch{\ast1} and
J.F. V\'azquez-Poritz\hoch{\dagger2}}

\begin{document}
\begin{flushright}

MIFP-03-20\ \ \ \ \
ULB-TH/03-27\\
{\bf hep-th/0308104}\\
August\  2003
\end{flushright}


\begin{center}

{\large {\bf Four-Dimensional Einstein Yang-Mills De Sitter
Gravity From Eleven Dimensions}}

\vspace{20pt}
\auth

\vspace{20pt} {\hoch{\ast}\it George P. and Cynthia W. Mitchell
Institute for Fundamental Physics,\\ Texas A\& M University,
College Station, TX 77843-4242, USA}

\vspace{10pt} {\hoch{\dagger}\brussels}

\vspace{30pt}

\underline{ABSTRACT}
\end{center}

       We obtain $D=4$ de Sitter gravity coupled to $SU(2)$ Yang-Mills
gauge fields from an explicit and consistent truncation of $D=11$
supergravity {\it via} Kaluza-Klein dimensional reduction on a
non-compact space. The ``internal'' space is a smooth hyperbolic
7-space ($H^7$) written as a foliation of two 3-spheres, on which the
$SU(2)$ Yang-Mills fields reside.  The positive cosmological constant
is completely fixed by the $SU(2)$ gauge coupling constant.  The
explicit reduction ansatz enables us to lift any of the $D=4$
solutions to $D=11$.  In particular, we obtain dS$_2$ in M-theory,
where the nine-dimensional transverse space is an $H^7$ bundle over
$S^2$. We also obtain a new smooth embedding of dS$_3$ in $D=6$
supergravity.

{\vfill\leftline{}\vfill \vskip 10pt \footnoterule {\footnotesize
\hoch{1} Research supported in part by DOE grant
DE-FG03-95ER40917.

{\footnotesize \hoch{2} Research supported in part by the Francqui
Foundation (Belgium), the Actions de Recherche \phantom{of the}
Concert{\'e}es of the Direction de la Recherche Scientifique -
Communaut\'e Francaise de Belgique, \phantom{of the} IISN-Belgium
(convention 4.4505.86) and by a ``Pole d'Attraction
Interuniversitaire.''}} \vskip -12pt}  \pagebreak
\setcounter{page}{1}

\newpage

\section{Introduction}

        The embedding of Anti-de Sitter (AdS) spacetimes in M-theory
and string theories is rather straightforward. In fact, gauged
supergravities in diverse dimensions with AdS vacuum have either been
shown or are expected to be obtainable from consistent Kaluza-Klein
sphere reductions of $D=11$ or $D=10$ supergravities.  Notable
examples include the simple embedding of $D=4$, ${\cal N}=2$, $SU(2)$
Yang-Mills AdS supergravity in $D=11$ \cite{pope1}, the significantly
more complicated $S^7$ \cite{dew,clpd4} and $S^4$
\cite{nas1,lpd7,nas2} reductions of M-theory, and the warped $S^4$
reduction of massive type IIA theory \cite{clpd6}.  Although the $S^5$
reduction of type IIB theory has yet to be fully
established,\footnote{The full metric ansatz was conjectured in
\cite{kpw}.} the reduction of a certain truncation of the theory has
been constructed \cite{lptd5,clpstd5}.

       On the other hand, there is less known regarding embedding de
Sitter (dS) spacetime in M-theory or string theories, mostly because
this is quite a bit more complicated than the case of AdS.  With
recent experimental evidence suggesting that our universe might be de
Sitter \cite{exp1,exp2}, there is increasing interest in de Sitter
gravity in cosmology and the dS/CFT correspondence
\cite{park1,park2,strominger1,strominger2,vijay}. Thus, it is of
importance to obtain the embedding of a non-trivial de Sitter gravity
theory in M-theory or string theories.

    The first de Sitter solution within the context of an extended
supergravity theory was found in \cite{gates}. While no-go
theorems \cite{gibbons,nogo} imply that de Sitter spacetime cannot
arise from a compactification of a supergravity theory, it can
arise from a supergravity theory with a non-compact ``internal''
space \cite{warner}. Explicit embeddings of dS$_4$ and dS$_5$ in
M-theory and type IIB supergravity, respectively, were obtained in
\cite{gh}. These arise as ten or eleven-dimensional solutions that
have a non-compact hyperbolic internal space.

    In this paper, we obtain four-dimensional de Sitter gravity with
$SU(2)$ Yang-Mills gauge fields from a less constrained truncation of
$D=11$ supergravity {\it via} Kaluza-Klein dimensional reduction on
the non-compact space.  In this construction, a consistent truncation
of the higher-dimensional theory is required in which there are no
modes which depend on the internal space.  The $SU(2)$ fields arise
from modes on the $S^3$ portions of the internal space. The Yang-Mills
gauge coupling constant is completely fixed by the cosmological
constant.  The kinetic terms for the $SU(2)$ gauge fields have the
correct sign, implying that the theory is not merely an analytical
continuation of $SU(2)$ AdS supergravity.

    This paper is organized as follows.  In section 2, we rederive the
embedding of dS$_4$ spacetime in M-theory, with the transverse
space being an $H^7$ written as a foliation of two 3-spheres.  In
section 3, we propose a reduction ansatz for obtaining $D=4$
$SU(2)$ Yang-Mills de Sitter gravity from $D=11$. We show that the
reduction ansatz is indeed consistent with the $D=11$ equations of
motion, and hence obtain the Lagrangian for $D=4$ $SU(2)$
Yang-Mills de Sitter gravity. The consistency of the reduction
ansatz enables us to lift any $D=4$ solution back to $D=11$.  In
section 4, we discuss this in detail.  In particular, we embed
dS$_2$ and (Minkowski)$_2$ spacetimes in $D=11$. The internal
space is an $H^7$ bundle over $S^2$. We also embed a cosmological
solution which smoothly interpolates between dS$_2\times S^2$ at
the infinite past in the co-moving time to a dS$_4$-type geometry
at the infinite future. In section 5, we obtain an embedding of
dS$_3$ in $D=6$ supergravity with the transverse space being an
$H^3$ written as a foliation of two circles. We conclude our paper
in section 6.

\section{Embedding dS$_4$ in $D=11$}

     We now show how the dS$_4$ embedding in $D=11$ supergravity found
in \cite{gh} can be obtained directly from the eleven-dimensional
equations of motion. We start with the Lagrangian of the bosonic
sector of $D=11$ supergravity, given by
\be {\cal L} = R\,{*\oneone} -\ft12 {*F_\4}\wedge F_\4 - \ft16
A_\3\wedge F_\4\wedge F_\4\,, \ee
where $F_\4=dA_\3$.  We consider the ansatz
\bea
ds^2 &=& H^2\, ds_4^2 + d\rho^2 + a^2\, d\Omega_3^2 + b^2\,
d\wtd\Omega_3^2\,,\nn\\
F_4&=&q\,\epsilon_\4\,,
\eea
where $H$, $a$ and $b$ are functions of $\rho$, $ds_4^2$ is
four-dimensional de Sitter spacetime with cosmological constant
$\Lambda=6\lambda^2$, {\it i.e.} $R_{\mu\nu}= 3 \lambda^2\,
g_{\mu\nu}$, and $\epsilon_\4$ is the corresponding volume-form.
$d\Omega_3^2$ and $d\wtd \Omega_3^2$ are the metrics of the two unit
3-spheres. The Einstein equations of motion are given by
\bea
\fft{4\ddot H}{H} + \fft{3\ddot a}{a} + \fft{3\ddot b}{b}&=&-
\fft{q^2}{6H^8}\,,\nn\\
\fft{\ddot H}{H} + \fft{3\dot H^2}{H^2} + \fft{3\dot H}{H}\,
\Big(\fft{\dot a}{a} + \fft{\dot b}{b}\Big) &=& \fft{q^2}{3H^8} +
\fft{3\lambda^2}{H^2}\,,\nn\\
\fft{\ddot a}{a} + \fft{2\dot a^2}{a^2} +
\fft{3\dot a\,\dot b}{a\,b} + \fft{4\dot a\,\dot H}{a\,H}
-\fft{2}{a^2} &=&-\fft{q^2}{6 H^8}\,,\nn\\
\fft{\ddot b}{b} + \fft{2\dot b^2}{b^2} +
\fft{3\dot a\,\dot b}{a\,b} + \fft{4\dot b\,\dot H}{b\,H}
-\fft{2}{b^2} &=&-\fft{q^2}{6 H^8}\,,\label{eom1}
\eea
where a dot represents a derivative with respect to $\rho$.  If the
metric $ds_4^2$ is AdS instead of dS, then one can have a solution
with $H$ being a constant. In this case, $\rho$ becomes an angular
coordinate with $a\sim\cos\rho$ and $b\sim \sin\rho$, so that the
metric becomes the direct product of AdS$_4$ and $S^7$, with the seven
sphere written as a foliation of two three spheres. Inspired by the
sphere reduction ansatz \cite{clpd4}, we consider the following
redefinition of variables
\be
H^2=\Delta^{2/3}\,,\qquad a^2= \Delta^{-1/3}\, \td a^2\,,\qquad
b^2=\Delta^{-1/3}\, \td b^2\,.\label{red1}
\ee
Following the analogous relation in the case of AdS$_4\times S^7$, we
have
\be
\td a^2 = \ft12 \ell^2\, (\Delta +1)\,,\qquad
\td b^2 = \ft12 \ell^2\, (\Delta -1)\,,\label{con1}
\ee
where $\ell$ is a constant scale parameter.  Substituting (\ref{red1})
and (\ref{con1}) into (\ref{eom1}), we find that the constants $q$ and
$\lambda$ must satisfy
\be
q^2\,\ell^2=4\,,\qquad \lambda^2\, \ell^2=\ft43\,.
\ee
The equations (\ref{eom1}) reduce to a single first-order
differential equation
\be
\ell^2\, \Delta^{\ft23}\, \dot \Delta^2 - 4\Delta^2 + 4=0\,.
\ee
Making a coordinate change $d\rho=\ell\, \Delta^{1/3}\,d\theta$,
we can easily solve for $\Delta$, which is given by
\be \Delta=\cosh\, (2\theta)\,.\label{delta} \ee
Thus we have an explicit embedding of dS$_4$ in $D=11$ given by
\bea
ds^2 &=& \cosh^{\ft23}(2\theta)\, ds_4^2 +
\ell^2\,\cosh^{\ft23}(2\theta)\, d\theta^2 \nn\\
&&+ \cosh^{-\ft13}(2\theta)\, \Big(\cosh^2\theta\,d\Omega_3^2
+\sinh^2\theta\,d\wtd \Omega_3^2\Big)\,,\nn\\
F_4 &=& \fft2{\ell}\,\epsilon_\4\,.\label{ds4ind11}
\eea
This solution was obtained in \cite{gh}.

Each $S^3$ in (\ref{ds4ind11}) can be replaced by a
three-dimensional lens space. A Kaluza-Klein reduction and Hopf
T-duality transformation on the fibre coordinate of the two lens
spaces yields an embedding of dS$_4$ in type IIB theory. This
procedure was used for the case of AdS solutions in \cite{warped}.

    If we consider (\ref{ds4ind11}) as a reduction ansatz from $D=11$
to $D=4$, then the resulting four-dimensional theory is Einstein
gravity with a positive cosmological constant, and with a
corresponding Lagrangian given by
\be
e^{-1}{\cal L} = R- \fft{8}{\ell^2}\,.
\ee

\section{Embedding Yang-Mills de Sitter gravity}

      The metric of the 3-spheres in (\ref{ds4ind11}) can be written as
\be
d\Omega_3^2 = \ft14 (\sigma_1^2 + \sigma_2^2 + \sigma_3^2)\,,\qquad
d\wtd \Omega_3^2 = \ft14 (\td\sigma_1^2 + \td \sigma_2^2 + \td
\sigma_3^2)\,,
\ee
where $\sigma_i$ and $\td \sigma_i$ are $SU(2)$ left-invariant
1-forms satisfying
\be
d\sigma_i =-\ft12 \epsilon_{ijk}\,\sigma_j\wedge\sigma_k\,,\qquad
d\td\sigma_i = -\ft12 \epsilon_{ijk}\,
\td\sigma_j\wedge\td\sigma_k\,.
\ee
Thus, we can introduce $SU(2)$ Yang-Mills fields $A_\1^i$ to the
vielbein
\be
h^i = \sigma_i - g\, A_\1^i\,,\qquad
\td h^i=\td \sigma_i - g\, A_\1^i\,.
\ee
With these preliminaries, we propose the reduction ansatz
\bea
ds_{11}^2 &=& \Delta^{\ft23}\, ds_4^2 + g^{-2}\,
\Delta^{\ft23}\, d\theta^2+ \ft14 g^{-2}\,
\Delta^{-\ft13}\, \Big[
c^2\, \sum_i(h^i)^2 + s^2\,\sum_i(\td h^i)^2\Big]
\,,\label{metred}\\
F_\4&=&2g\, \epsilon_\4 -\ft14 g^{-2} \Big(s\,c\, \, d\theta\wedge
h^i\wedge {*F_\2^i} -s\,c\, \, d\theta \wedge \td h^i\wedge
{*F_\2^i}\nn\\
&&- \ft14c^2\, \epsilon_{ijk}\,h^i\wedge h^j\wedge {*F_\2^k} +
\ft14s^2\, \epsilon_{ijk}\, \td h^i\wedge\td h^j\wedge
{*F_\2^k}\Big)\,, \label{f4red} \eea
where $c=\cosh \theta$, $s=\sinh \theta$, $\Delta=\cosh(2\theta)$, and
* denotes the four-dimensional Hodge dual.  Note that we have
rewritten the scale parameter $\ell$ of section 2 in terms of
$g=\ell^{-1}$. The $SU(2)$ Yang-Mills field strengths $F_\2^i$ are
given by
\be
F_\2^i=dA_\1^i + \ft12 g\, \epsilon_{ijk} A_\1^j\wedge A_\1^k\,.
\ee
The $D=11$ Hodge dual of the 4-form is given by
\bea
{\hat *F_\4} &=& \ft1{32}\, g^{-6}\, \Delta^{-2}\, c^3\,s^3\,
d\theta\wedge\epsilon_\3\wedge \td\epsilon_\3\nn\\
&&-\ft1{128}\,g^{-5}\,\Delta^{-1}\, c^2\, s^4\, \epsilon_{ijk}
\,h^i\wedge h^j\wedge F_\2^k\wedge\td\epsilon_\3\nn\\
&&-\ft1{128}\,g^{-5}\,\Delta^{-1}\, s^2\, c^4\,
\epsilon_{ijk}\,\td h^i\wedge \td h^j\wedge F_\2^i\wedge
\epsilon_\3\nn\\
&&+\ft1{32}\, g^{-5}\,c\,s^3\, d\theta\wedge h^i\wedge
F_\2^i\wedge \td\epsilon_\3 +
\ft1{32}\,g^{-5}\,s\,c^3\,d\theta\wedge \td h^i\wedge F_\2^i
\wedge \epsilon_\3\,. \eea

     It is straightforward to verify that the Bianchi identity
$dF_\4=0$ is satisfied provided that the $SU(2)$ Yang-Mills fields
$A_\2^i$ satisfy the lower-dimensional equations of motion
\be
D{*F_\2^i}=0\,,
\ee
where the covariant derivative $D$ is defined by $DV^i=dV^i + g\,
\epsilon_{ijk}\, A^j\wedge V^k$, for any vector $V^i$.  The following
identities are useful in verifying the equations of motion
\be
DF_\2^i=0\,,\quad Dh^i=-\ft12\epsilon_{ijk}\,h^j\wedge h^k-
g\,F_\2^i\,\quad D\td h^i=-\ft12\epsilon_{ijk}\,\td h^j\wedge
\td h^k- g\, F_\2^i\,.
\ee
The following formulae are also useful
\bea
d(h^i\wedge {*F_\2^i}) &=& Dh^i\wedge {*F_\2^i} - h^i\wedge D{*F_\2^i}
\nn\\
&=&-\ft12 \epsilon_{ijk}\,h^j\wedge h^k\wedge {*F_\2^i}
- g\, F_\2^i \wedge {*F_\2^i}\,,\nn\\
\epsilon_{ijk}\,d(h^i\wedge h^j\wedge {*F_\2^k})&=&
\epsilon_{ijk}\, D(h^i\wedge h^j)\wedge {*F_\2^k}=0\,.
\eea
The verification of $d{\hat *F_\4}=\ft12 F_4\wedge F_\4$ requires the
following identity
\bea
\Delta^{-2}\,c^3\,s^3 -\ft12 (\Delta^{-1}\, c^2\, s^4)' +
c\,s^3 &=&0\,,\nn\\
-\Delta^{-2}\,c^3\,s^3 -\ft12 (\Delta^{-1}\, s^2\, c^4)' +
s\,c^3 &=&0\,.
\eea

      The evaluation of the $D=11$ Einstein equations of motion are
much more complicated, and we have not performed the calculation
in full detail.  However, we have verified that the equations of
motion work for the $U(1)$ subsector of the $SU(2)$ gauge fields.
Combining the result, the lower-dimensional equations of motion
can be obtained from the Lagrangian
\be
e^{-1} {\cal L} = R - \ft14 (F_\2^i)^2 - 8 g^2\,.\label{d4lag}
\ee
The cosmological constant $8g^2$ is totally fixed by the gauge
coupling constant $g$.  Thus, we have obtained four-dimensional
Einstein $SU(2)$ Yang-Mills de Sitter gauged gravity from $D=11$ by
consistent Kaluza-Klein reduction on a hyperbolic 7-space.

        It is worth remarking that $D=11$ supergravity can also give
rise to $D=4$ $SU(2)$ AdS supergravity \cite{pope1}. Also, de
Sitter gravity with the wrong sign in the kinetic terms for gauge
fields can arise from hyperbolic reduction of * variations of
M-theory, type IIB or massive type IIA theories
\cite{lsw,lvpdstods}. In our reduction of M-theory, however, the
sign of the kinetic term for $A_\1^i$ is the right one.

\section{Lifting of solutions}

    The four-dimensional Lagrangian (\ref{d4lag}) admits a large class
of solutions, including multi-center black holes
\cite{ls,behrndt,lsw}. Using our reduction ansatz (\ref{metred})
and (\ref{f4red}), all of these solutions can be lifted to eleven
dimensions. We will first explicitly lift the case of dS$_2\times
S^2$, which is supported by one of the $SU(2)$ gauge fields, {\it
e.g.}~$F_\2^3$. This solution is given by
\bea
ds_4^2 &=& -d\tau^2 + e^{\ft{\tau}{\ell\,\sqrt2}}\, dx^2 +
4\ell^2\, d\Omega_2^2\,,\nn\\
F_\2^3 &=& \ft{1}{2\ell}\,e^{\ft{\tau}{2\sqrt2\,\ell}}\,
d\tau\wedge dx\,,\qquad {*F_\2^3} = 2\ell\, \Omega_\2\,,
\label{d4ds2}
\eea
where $\ell^2=3/(64g^2)$.  Note that the role of $F_\2^3$ and
${*F_\2^3}$ are interchangeable, giving rise to either the electric or
magnetic solutions, with the metric unchanged. Lifting this solution
back to $D=11$ yields a smooth and regular embedding of dS$_2$ given
by
\bea
ds_{11}^2 &=& \Delta^{\ft23}\, \Big(-d\tau^2 +
e^{\ft{\tau}{\ell\,\sqrt2}}\, dx^2 + 4\ell^2\, d\Omega_2^2
\Big) + g^{-2}\,\Delta^{\ft23}\, d\theta^2\nn\\
&&+\ft14g^{-2}\, \Delta^{-\ft13}\, \Big[
c^2\,\Big(d\omega_2^2 + (\sigma_3 -
\sqrt2\,g\,e^{\ft{\tau}{2\sqrt2\,\ell}}\,dx)^2\Big)\nn\\
&&\qquad\qquad\quad
 + s^2\,\Big(d\wtd\omega_2^2 + (\td\sigma_3 -
\sqrt2\,g\, e^{\ft{\tau}{2\sqrt2\,\ell}}\,dx)^2\Big)\Big]\,,\nn\\
F_\4 &=& 8g\,\ell^2\, e^{\ft{\tau}{2\sqrt2\,\ell}}\,d\tau\wedge
dx\wedge\Omega_\2\\
&&-\ft12g^{-2}\,\ell\, \Big(s\,\, c\,\,
d\theta\wedge(\sigma_3-\td\sigma_3)
 - \ft12c^2\, \omega_\2\wedge +
\ft12s^2\,\td\omega_\2\Big)\wedge \Omega_\2\,.\nn
\eea
We explicitly verified that the above solution satisfies the equations
of motion of $D=11$ supergravity, which serves as a double check of
our reduction ansatz (\ref{metred}) and (\ref{f4red}). In this smooth
embedding of dS$_2$ in M-theory, the metric can be viewed as a
rotating brane. Of course, if we interchange the role of $F_\2^3$ and
$*F_\2^3$ in (\ref{d4ds2}), then the transverse space is a
nine-dimensional non-compact space which can be viewed as an $H^7$
bundle over $S^2$.

We will now consider a regular cosmological solution of
(\ref{d4lag}) given by
\bea ds_4^2 &=& H^2\, (-f^{-1}\, dt^2 + f\, dx^2 +
t^2\, d\Omega_2^2)\nn\\
F_\2^3 &=& \fft{2\ell}{(t\,H)^2}\, dt\wedge dx\,,\qquad
{*F_\2^3}=2\ell\, \, \Omega_\2\,,\nn\\
H&=&1 +\fft{\ell}{t}\,,\qquad f=\ft43g^2 t^2\, H^4 -1\,. \eea

This solution is, in fact, nothing but the BPS AdS
Reissner-Nordstr\o m black hole with $g \rightarrow \im\, g$
\cite{adsbh}.  When $g^2\,\ell^2 = \ft{3}{64}$, the solution
interpolates between dS$_2\times S^2$ at the infinite past in the
co-moving time to a dS$_4$-type geometry at the infinite future
with the boundary of $S^2\times S^1$ \cite{adsbh}. It is
straightforward to lift the solution back to $D=11$ and obtain a
regular cosmological solution in M-theory.  The corresponding
metric is given by
\bea ds_{11}^2 &=& \Delta^{\ft23}\, H^2\, \Big( -f^{-1}\, dt^2 +
f\, dx^2 + t^2\, d\Omega_2^2 \Big) +
g^{-2}\,\Delta^{\ft23}\, d\theta^2\\
&&+\ft14g^{-2}\, \Delta^{-\ft13}\, \Big[ c^2\,\Big(d\omega_2^2 +
(\sigma_3 -\fft{2g}{H}\,dx)^2\Big) + s^2\,\Big(d\wtd\omega_2^2 +
(\td\sigma_3 -\fft{2g}{H}\,dx)^2\Big)\Big]\,.\nn \eea

As a final example, we will lift (Minkowski)$_2\times S^2$ to
eleven dimensions. This four-dimensional solution is given by
\bea ds_4^2 &=& -d\tau^2 + dx^2 +
\fft{1}{8g^2}\, d\Omega_2^2\,,\nn\\
F_\2^3 &=& 4g\, \, d\tau\wedge dx\,,\qquad {*F_\2^3} =
\fft{1}{2g}\, \Omega_\2\,. \label{d4m2} \eea
Lifting this solution back to $D=11$ yields a smooth and regular
embedding of M$_2$, whose metric is given by
\bea ds_{11}^2 &=& \Delta^{\ft23}\, \Big(-d\tau^2 + dx^2 +
\fft{1}{8g^2}\, d\Omega_2^2
\Big) + g^{-2}\,\Delta^{\ft23}\, d\theta^2\nn\\
&&+\ft14g^{-2}\, \Delta^{-\ft13}\, \Big[ c^2\,\Big(d\omega_2^2 +
(\sigma_3 -
4g^2\,dx)^2\Big)\nn\\
&&\qquad\qquad\quad
 + s^2\,\Big(d\wtd\omega_2^2 + (\td\sigma_3 -
4g^2\,dx)^2\Big)\Big]\,. \eea
As in the previous examples, one can interchange the role of
$F_\2^3$ and $*F_\2^3$ in (\ref{d4m2}).

\section{Further embeddings of dS spacetime}

    In \cite{gh}, there are further embeddings of dS spacetime in
M-theory or type IIB supergravities.  In our example of a dS$_4$
embedding in M-theory, the $H^7$ is a foliation of $S^3\times
S^3$. The embeddings of dS$_4$ in M-theory with a squashed $H^7$ being
a foliation of $S^2\times S^4$ and dS$_5$ in type IIB theory with the
$H^5$ being a foliation of $S^2\times S^2$ were obtained in
\cite{gh}. In this section, we obtain an new embedding of dS$_3$ in
$D=6$ supergravity. The relevant Lagrangian is given by
\be
e^{-1}\,{\cal L} =  R-\ft12(\del\phi)^2 -
\ft1{12} e^{\sqrt2\phi}\,F_\3^2\,.
\ee
The solution is given by
\bea
ds_6^2 &=& \Delta\, ds_3^2 + g^{-2}\, \Delta\, d\theta^2 +
g^{-2}\,\Delta^{-1}\, (c^2\, d\phi_1^2 + s^2\, d\phi_2^2)\,,\nn\\
F_\3 &=& 2g\,(\epsilon_\3 + \    *\ psilon_\3)\,,\qquad
\phi=0\,,
\eea
where $ds_3^2$ is dS$_3$ spacetime with cosmological constant
$\Lambda=2g^2$.  Note that the transverse space is an $H^3$ written as
a foliation of two circles.  This theory can be reduced to pure de
Sitter gravity in three dimensions, with the corresponding Lagrangian
given by
\be
e^{-1}{\cal L}_3= R - 2g^2\,.
\ee

\section{Conclusions}

      We have obtained $D=4$ de Sitter gravity coupled to $SU(2)$
Yang-Mills fields from a consistent Kaluza-Klein reduction and
truncation of $D=11$ supergravity on a hyperbolic 7-space.  The
hyperbolic space is written as a foliation of two 3-spheres, on
which the $SU(2)$ fields are embedded.  Unlike in the case of *
theories, the kinetic terms of the gauge fields have the correct
sign.  The four-dimensional cosmological constant is completely
fixed by the gauge coupling constant. Although our reduction
procedure fits within the general pattern of non-compact gaugings
and their higher-dimensional origins, described in \cite{warner},
our result provides the first explicit embedding of a non-trivial
de Sitter gauge gravity in M-theory.

    The reduction ansatz enables us to lift any solution of the
four-dimensional theory to eleven dimensions.  We discuss the
embeddings of smooth cosmological solutions.  In particular, we
obtain the embeddings of dS$_2$ and M$_2$ in M-theory, as well as
that of a cosmological solution which smoothly interpolates
between dS$_2\times S^2$ at the infinite past in the co-moving
time to a dS$_4$-type geometry at the infinite future.  We also
obtain a new embedding of dS$_3$ in $D=6$ supergravity. Our
results provide important tools with which to study the dS/CFT
correspondence from the point of view of string and M-theory.

\section*{Acknowledgment}

       We are grateful to Chris Pope for useful discussions.

\end{document}